\documentclass[aps,prl,twocolumn,showpacs]{revtex4} 
\usepackage{float,graphicx}
\usepackage{dcolumn}
\begin{document}
\title{Small Fermi energy and phonon anharmonicity in
MgB$_2$ and related compounds}
\author{L. Boeri$^1$, G.B. Bachelet$^{1,2}$, E. Cappelluti$^1$,
L. Pietronero$^{1,2,3}$} 
\affiliation{$^1$INFM Unit\`a di Roma 1 and Dipartimento di Fisica, 
Universit\`a di Roma ``La Sapienza'', P.le Aldo Moro 2, 00185 Roma, 
Italia}
\affiliation{$^2$INFM Center for Statistical Mechanics and Complexity,
P.le Aldo Moro 2, 00185 Roma, Italia}
\affiliation{$^3$CNR, Istituto di Acustica ``O.M. Corbino'', v. del Fosso 
del Cavaliere 100, 00133 Roma, Italy}
\date{\today}
\begin{abstract}
The remarkable anharmonicity of the $E_{2g}$ phonon in MgB$_2$ has
been suggested in literature to play a primary role in its
superconducting pairing.  We investigate, by means of LDA 
calculations, the microscopic origin of such an anharmonicity in 
MgB$_2$, AlB$_2$, and in heavily hole--doped graphite.  We find 
that the anharmonic character of the $E_{2g}$ phonon is essentially 
driven by the small Fermi energy of the $\sigma$ holes.  We present 
a simple analytic model which allows us to understand in 
microscopic terms the role of the small Fermi energy and of the 
electronic structure.  The relation between anharmonicity and 
nonadiabaticity is pointed out and discussed in relation to various 
materials.
\end{abstract}
\pacs{74.70.Ad, 63.20.Ry, 63.20.Kr}
\maketitle
\begin{figure}[H]
\centering
\includegraphics[width=8cm]{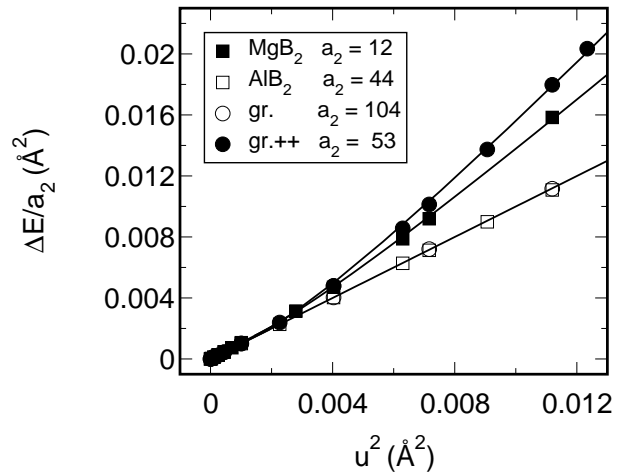}
\vspace{-4pt} \caption{\label{etot} Energy $\Delta E$ associated to 
an $E_{2g}$ phonon displacement of amplitude $u$, plotted as a 
function of $u^2$.  For each material this energy is divided by 
$a_2$ (in the inset, units of eV/\AA$^2$), the quadratic 
coefficient of a polynomial best fit $\Delta E \simeq a_2 u^2 + a_4 
u^4 + ...$.  On both axes the units are thus \AA$^2$, and harmonic 
phonons collapse on a single straight line $y=x$.  The solid lines 
result from our model (Eqs.~\ref{ensmall} and \ref{enlarge}); the 
corresponding parameters are shown in Table \ref{table1}.}
\end{figure}

 
The report of superconductivity at $T_c\!=\!39$ K in MgB$_2$ 
\cite{akimitsu} has raised great expectations about metal diborides 
$M$B$_2$.  There is indeed no reason to believe that MgB$_2$ 
represents the highest--$T_c$ compound within this family.  Since 
the very beginning a generic consensus about the electron-phonon 
(e--ph) nature of the superconducting pairing has prevailed 
\cite{bohnen,kong}, although some purely electronic models have 
been proposed\cite{hirsch}.  The precise origin of such a high--
$T_c$ superconducting phase is still unknown.
Recently the strong anharmonic character of the in-plane $E_{2g}$ 
phonon mode and its possible correlation with the high $T_c$ value 
in MgB$_2$ have attracted a considerable interest 
\cite{yildirim,liu,kunc,struzhkin,choi}.  Here we attempt a simple 
theory of such a strong anharmonicity and test its predictive power 
on related compounds.  With this perspective we have performed 
first--principles calculations of the band structure and lattice 
properties of MgB$_2$, AlB$_2$, and of a hypothetical hole--doped 
graphite.  We identify the small value of the Fermi energy for the 
holes in the $\sigma$ band, with entire portions of the Fermi 
surface disappearing upon $E_{2g}$ distortion \cite{liu,meletov}, 
as the fundamental origin of anharmonicity; a simple way of 
modeling the effect of distortion on the band structure confirms 
our finding.
It has already been pointed out in literature that MgB$_2$ 
resembles 
in many ways graphite \cite{kong,kortus,Satta,Pickett}.  From the structural point 
of
view MgB$_2$ is formed by graphene--like layers of B spaced by 
planes of Mg atoms.  The point group symmetry of in--plane boron 
phonon modes of MgB$_2$ and in--plane phonons of graphite is thus 
the same \cite{kunc}, and the difference in frequency is related to
the different strength of B--B and C--C bonds.  Out of the whole 
vibrational spectrum, a large interest has converged towards the 
$E_{2g}$ mode in MgB$_2$, which involves only in--plane boron 
displacements.  This mode has been shown to have an extremely 
strong coupling with the in--plane $\sigma$ bands 
\cite{yildirim,bohnen,kong,liu,kunc,choi,struzhkin}, which in 
MgB$_2$ provide conduction holes; a relation to the high 
superconducting temperature $T_c$ was naturally suggested.  This 
idea was also supported by the negligible partial isotope 
coefficient on $T_c$ associated with the Mg atomic mass 
\cite{hinks}.  In this context the strong anharmonicity, a unique 
property of the $E_{2g}$ phonon within 
the MgB$_2$ vibrational spectrum \cite{yildirim,kunc}, acquires an 
obvious importance.
The electronic structure of MgB$_2$ shares strong similarities with 
graphite.  In both materials one can identify strongly
two--dimensional $\sigma$ bands, almost entirely derived from B (C) 
$s$ and $p_{x,y}$ orbitals, plus bonding and antibonding $\pi$ 
bands with three--dimensional dispersion and mainly B$_{p_z}\!\!$--
Mg (C$_{p_z}$) character.  The most important difference between 
MgB$_2$ and graphite is the position of the Fermi level $\mu$.  In 
undoped graphite $\mu$ cuts the $\pi$ band structure just in the 
middle, about 3 eV above $\epsilon_\sigma^{\rm top}$, the top of 
the bonding $\sigma$ bands, which are thus completely full 
\cite{dresselhaus}. In MgB$_2$ the combined effect of additional 
magnesium layers, different ionic charge between boron and carbon, 
and valence--charge transfer from magnesium to boron layers, yields 
a different arrangement of $\sigma$ and $\pi$ bands.  The resulting 
Fermi level $\mu$ still cuts the $\pi$ band structure somewhere, 
but is now $\sim$0.5 eV below the top of the $\sigma$ bands, which 
therefore, in MgB$_2$, give a sizable hole contribution to the 
Fermi surface \cite{Pickett,kortus}, not present in graphite.  This 
important difference was almost immediately pointed out 
\cite{Pickett}; more recently the large splitting and shifts which 
the $\sigma$ bands undergo upon typical $E_{2g}$ phonon 
displacements were suggested as the likely source of anharmonicity 
for the $E_{2g}$ phonon in MgB$_2$ \cite{yildirim}.  We are going 
to make this statement more precise. We claim that neither the 
presence at the Fermi level of the $\sigma$ bands nor their strong 
coupling to the $E_{2g}$ phonon are sufficient to induce anharmonic 
effects: it's the small Fermi energy associated, in the unperturbed 
crystal, to the $\sigma$ conduction holes ($\epsilon_\sigma^{\rm 
top} - \mu \simeq 0.45$ eV) which makes the difference.  On these 
grounds we may expect anharmonic effects to be strong in other 
materials with small Fermi energies (and sufficient e--ph 
coupling), and also conjecture a relation between anharmonic and 
nonadiabatic effects.
We present local Density Functional \cite{HKS} calculations, based 
on Martins-Troullier pseudopotentials \cite{MartTroul} and the 
ABINIT code \cite{ABINIT}, which support this picture; the 
resulting bands and $E_{2g}$ frozen--phonon energies are well 
understood in terms of a simple model, discussed in the second part 
of this letter.  We studied the $E_{2g}$ phonon for MgB$_2$, 
AlB$_2$, graphite and a hypothetical hole--doped graphite where one 
electron is missing from each carbon atom.  Experimental lattice 
parameters were used as an input for MgB$_2$, AlB$_2$ and graphite, 
while for hole--doped graphite the lattice parameters were left at 
the experimental value of graphite. For the purpose of this study 
the use of GGA's and an extremely accurate {\it a priori} 
determination of equilibrium lattice parameters are not an issue.  
Great care in the {\bf k}-space integration is, instead, an issue 
\cite{kunc}, since, in some cases, entire portions of the Fermi 
surface disappear upon distortion; we
use two $15\!\times\!15\!\times\!10$ shifted Monkhorst-Pack 
grids in the Brillouin zone \cite{ABINIT}.  We should also 
specify that, 
between the two possible eigenvectors for the $E_{2g}$ phonon, we 
only show results for the one labeled $E_{2g}(b)$ in 
Ref.\cite{kunc}, whose energy is, by symmetry, an even function of 
the displacement; phonon displacements up to $0.05 \sim 0.1$ \AA \, 
were considered.
The first important observation is that the $E_{2g}$ anharmonicity 
is completely absent in AlB$_2$, whose $\sigma$ bands undergo 
equally large splittings and shifts as MgB$_2$.  As shown in 
Fig.~\ref{etot}, our frozen--$E_{2g}$--phonon calculations 
reproduce the large anharmonicity found for MgB$_2$ (black squares) 
\cite{yildirim,kunc}, but predict no such effect for AlB$_2$ (empty 
squares), whose energy remains proportional to the square of the 
phonon displacement for all the displacements under consideration.  
Our finding is consistent with the experimental fact that the 
$E_{2g}$ phonon line is very broad in MgB$_2$ but not in AlB$_2$ 
\cite{bohnen}.
\begin{figure}[H]
\centering
\includegraphics[width=8cm]{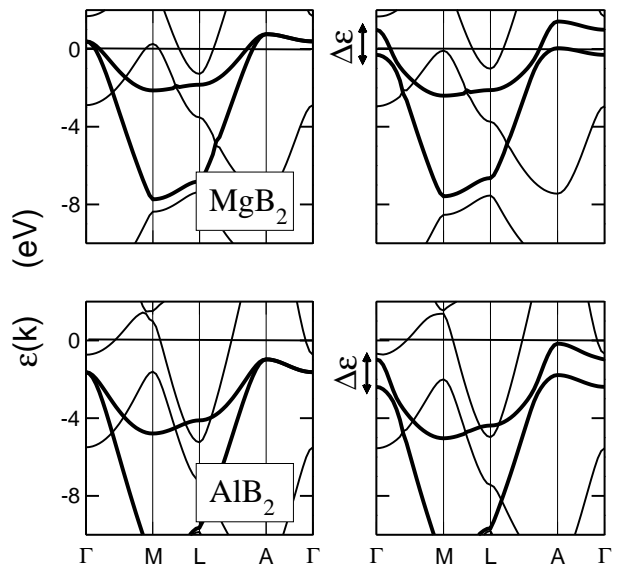}
\vspace{-4pt} \caption{\label{bands} Upper panels: MgB$_2$ 
electronic bands without (left) and with (right) an $E_{2g}$ 
phonon 
distortion of amplitude $u=$ 0.05 \AA. Lower panels: same as upper 
panels, but for AlB$_2$ and $u=$ 0.05$\times
({a_{AlB_{2}}}/{a_{MgB_{2}}})$ \AA. The $\sigma$ bands are marked as 
thicker lines.  Their splitting upon $E_{2g}$ distortion is equally 
large in MgB$_2$ and AlB$_2$, but in the latter (lower
panels) they are always below the Fermi level ($\mu\!=\!0$ in all 
panels).  }
\end{figure}
We suggest that the different behavior of AlB$_2$ be simply related to 
the fact that, both before and after the $E_{2g}$ distortion, its 
$\sigma$ bands are near, but completely below, the Fermi level $\mu$; 
unlike MgB$_2$, their electronic occupation remains unchanged upon 
distortion (see Fig.~\ref{bands}).  In MgB$_2$, instead, the top of
the $\sigma$ bands is above the Fermi energy but, upon distortion, 
the lower splitoff band completely sinks below it, thus changing 
its occupation.  Besides the amount of the shifts and splittings, 
the exact position of the $\sigma$ bands before and after the 
$E_{2g}$ distortion is thus a crucial ingredient for its 
anharmonicity.
This is confirmed by artificially moving the top of these bands, 
$\epsilon_{\sigma}^{top}$, w.r.t. the Fermi level $\mu$ in 
graphite. In true graphite the $\sigma$ bands also undergo large 
splittings, but they are already well below $\mu$ both before and 
after the $E_{2g}$
distortion; here our frozen--$E_{2g}$--phonon calculations find no 
anharmonicity (Fig.~\ref{etot}, empty dots; ``gr.''  stays for 
graphite).  But if, by adding a uniform neutralizing background 
\cite{Pickett}, we remove one electron per carbon atom (two 
electrons per cell, black dots in Fig.~\ref{etot}) from graphite, 
thus shifting $\epsilon_{\sigma}^{top}$, the top of its $\sigma$ 
bands, back to the ``optimal'' position ($\sim 1$ eV above $\mu$ in 
the undistorted crystal), then, upon typical $E_{2g}$ distortions, 
the lower splitoff $\sigma$ band sinks below $\mu$, and we recover 
strong anharmonic effects, shown in Fig.~\ref{etot} (black dots).
To clarify the origin of this anharmonic behavior we have traced 
back the effect of the $E_{2g}$ frozen--phonon distortion to the 
electronic structure.  For all materials we find that the main 
effect of the lattice displacement is a linear energy splitting of 
the $\sigma$ bands.
\begin{table}[h]\begin{center}
\begin{tabular}{c||c|c|c||c|c}
	&  g  & $\epsilon_{\sigma}^{top}$ & a$_2$ \\
\hline
\hline
MgB$_2$ &  12.02 & 0.45 & 12 \\
\hline
AlB$_2$ &  11.74 & -1.63 & 44  \\
\hline
gr.     &  28.29 & -2.89 & 104 \\
\hline
gr.++   &  30.86& 1.17 & 53 \\
\end{tabular}
\caption{\label{table1} 
Three inputs for our total energy model, Eqs.~(\ref{ensmall})
and (\ref{enlarge}), extracted from our LDA outputs.  The
remaining two parameters N$_{\sigma}$ and N$_{\pi}$, needed only 
when $\epsilon_{\sigma}^{top}\!>\!\mu$, were adjusted to yield the 
best fits shown in Fig.~\ref{etot}.  Their optimal values 
(N$_{\sigma}\!=\!0.11$, N$_{\pi}\!=\!0.39$ for MgB$_2$; 
N$_{\sigma}\!=\!0.07$, N$_{\pi}\!=\!0.30$ for graphite++) fall in a 
physically reasonable
range, in spite of our oversimplified density of states (see 
text).} \end{center}
\vspace{-20pt}
\end{table}
The $\pi$ bands are, instead, only weakly modified.  The effects of 
lattice distortion on the band structure, in a relevant energy 
range around the Fermi level $\mu$, can be thus schematized, to a 
good approximation, as $\delta\epsilon_\sigma({\bf k}) \simeq \pm g 
|u|$, with opposite signs for the two different $\sigma$ bands.  
The value of $g$ deduced from the LDA results differs significantly 
between the borides (MgB$_2$, AlB$_2$) and graphite (with or 
without doping), while it is almost constant within each class of 
compounds (see Table \ref{table1}).
>From the comparison of the band structures in Fig.~\ref{bands} we 
can identify two representative cases.  For AlB$_2$ (bottom) the 
bonding $\sigma$ bands, at zero distortion (left), are completely 
below the Fermi level $\mu$, and remain there upon distortion 
(right); their energy splitting, induced by the $E_{2g}$ lattice 
displacement, does not change either their occupation or the 
topology of the Fermi surface, which never acquires a $\sigma$ 
sheet; no $\sigma$ band crosses $\mu$ at any displacement, and the 
Fermi surface is exclusively dictated by the $\pi$ bands, which, 
compared to the $\sigma$ bands, undergo only minor changes upon 
displacement.  Besides
AlB$_2$, this is also the case of undoped graphite (not
shown).  An entirely different situation is found when, on one 
hand, the top of the bonding $\sigma$ bands at zero distortion, 
$\epsilon_{\sigma}^{top}$, is above the Fermi level $\mu$ (so that 
in the perfect crystal the Fermi surface has $\sigma$--hole--like 
cylindrical sheets \cite{kortus}), but, on the other, the energy 
splitting of the $\sigma$ bands is large enough to drive one of 
them completely below $\mu$ upon distortion.  This is the case 
of MgB$_2$ (top Fig.~\ref{bands}), and also of heavily hole--doped 
graphite (not shown).  In both cases, at some critical phonon 
displacement, the number of Fermi surfaces associated to the 
$\sigma$ bands changes (one of the two cylindrical sheets around 
the $\Gamma$--A line disappears \cite{liu}); beyond that point, 
since 
the total number of electrons is conserved, larger displacements 
will imply a qualitatively different behavior, due to the 
reshoveling of electrons between $\sigma$ and $\pi$ states.
To gain further insight, we present a simple model of the
$E_{2g}$ anharmonicity which seems to represent well both types of 
situations.  We first consider the system with no distorsion 
($u\!=\!0$).  The electronic band energy can be written as: 
\begin{equation}
E(u\!=\!0)=2\sum_{{\bf k},i}
\epsilon_i({\bf k}) n_i({\bf k})
+2\sum_{{\bf k}}
\epsilon_\pi({\bf k}) n_\pi({\bf k}),
\label{zerodist}
\end{equation}
where $\epsilon_i({\bf k})$ represents the dispersion relation of 
the two $\sigma$ bands, $\epsilon_\pi({\bf k})$ takes into account 
the remaining $\pi$ bands, and $n_i({\bf k})$, $n_\pi({\bf k})$ are 
the corresponding occupations (the factor 2 is for the spin 
degeneracy). The LDA bands teach us that the most important effect 
of an $E_{2g}$ distortion $u\!\neq\!0$ is an almost linear 
splitting (i.e. proportional to $u$) of the two $\sigma$ bands 
around the $\Gamma$--A line.  This can be roughly modeled by a 
linear e--ph Jahn--Teller--like coupling of the $E_{2g}$ phonon to 
the $\sigma$ bands; the small coupling to the $\pi$ bands is 
neglected altogether. The resulting total energy at $u\!\neq\!0$ 
is:
\begin{eqnarray}
E(u) = &2& \sum_{{\bf k},i}
\epsilon_i({\bf k}) n_i({\bf k},u)
+ 2\sum_{{\bf k}}
\epsilon_\pi({\bf k}) n_\pi({\bf k},u)
\nonumber\\
+ &2& \!\!\!g u \sum_{{\bf k}}
\left[
n_2({\bf k},u)-n_1({\bf k},u)
\right]
+ \frac{M \omega_{2g}^2}{2} \, u^2
\label{entot}
\end{eqnarray}
In Eq.~(\ref{entot}) the electronic band energy, in the presence 
of a phonon displacement $u$, was split into the sum of three 
terms: unperturbed $\sigma$ and $\pi$ bands with $u$--dependent 
occupation (first two terms in the r.h.s), plus a linear e--ph 
coupling of the $E_{2g}$ mode with the $\sigma$ bands.  The last 
term is an effective elastic energy.  Far from the Fermi level our 
bands are not realistic (see below), so we lump into this term 
both the bare ion--ion repulsion and those electronic effects 
which are missing from our model bands.  The occupation number can 
be self--consistently calculated: $n_1({\bf 
k},u)=f[\epsilon_1({\bf k})-gu-\mu(u)]$, $n_2({\bf 
k},u)=f[\epsilon_2({\bf k})+gu-\mu(u)]$, $n_\pi({\bf 
k},u)=f[\epsilon_\pi({\bf k})-\mu(u)]$, where $\mu(u)$ is the 
Fermi level in the presence of the frozen phonon and 
$f[x]=\theta(-x)$ is the $T\!=\!0$ Fermi function.

Let us consider the representative case of MgB$_2$.  For the sake 
of simplicity we assume two parabolic $\sigma$ bands, perfectly 
two--dimensional and degenerate at $u\!=\!0$ [$\epsilon_1({\bf 
k})\!=\!\epsilon_2({\bf k})\!=\!\epsilon_\sigma({\bf k})$].  The 
corresponding density of states (DOS) of each $\sigma$ band will be 
therefore constant up the top of the band $\epsilon_\sigma^{\rm 
top}$: $N_\sigma(\epsilon)\!=\!N_\sigma$ [$\epsilon \le 
\epsilon_\sigma^{\rm top}$].  From now on we conveniently set 
$\mu(u\!=\!0)=0$, so that $\epsilon_\sigma^{\rm top}$ now equals 
$\epsilon_\sigma^{\rm top}\!-\!\mu(u\!=\!0)$, the Fermi energy of 
the $\sigma$ holes in the absence of lattice distorsion.  In 
addition, in the energy range we are interested of, we can assume 
$N_\pi$, the density of states of the $\pi$ band, to be just 
constant.
The Fermi level $\mu(u)$ and the total energy $E(u)$ in the 
presence of the frozen phonon distorsion $u$ can now be easily 
computed.  The system shows a qualitatively different behavior for 
two interesting regimes: ($i$) $g |u| \le \epsilon_\sigma^{\rm 
top}$ and ($ii$) $g |u| \ge \epsilon_\sigma^{\rm top}$.  Within the 
assumptions of our model (rectangular DOS), in the regime ($i$) the 
Fermi level is unaffected by the frozen phonon distorsion, 
$\mu(u)\!=\!\mu(u\!=\!0)\!=\!0$: the depletion of the $\sigma$ 
band, raised by the distortion, is compensated by an equivalent 
filling of the other $\sigma$ band, lowered by the same amount; the 
$\pi$ bands, modeled by their constant DOS $N_{\pi}$, play no role.  
We obtain for the total energy: \begin{equation}
E(u)=E(0)+\frac{M\omega_{2g}^2}{2}u^2 - 2 N_\sigma g^2 u^2 
\hspace{5mm} g |u| \le \epsilon_\sigma^{\rm top}.
\label{ensmall}
\end{equation}
Eq.  (\ref{ensmall}) represents a phonon frequency renormalization 
due to the response of the $\sigma$ electrons, $E(u)=E(0)+ 
M\Omega_{2g}^2 u^2 / 2$, with $\Omega_{2g}^2=\omega_{2g}^2 - 4 
N_\sigma g^2/M$.  The harmonic character of the $E_{2g}$ phonon 
mode is however unaffected. Things change in the regime ($ii$) $g 
|u| \ge \epsilon_\sigma^{\rm top}$.  When the energy splitting is 
larger than the zero--distortion Fermi energy of the $\sigma$ 
holes, $g |u| \le \epsilon_\sigma^{\rm top}$, the lower $\sigma$ 
band is completely shifted below the Fermi level.  Now this band is 
full, and cannot further compensate the loss of electrons from the 
upper $\sigma$ band.  Then the only way to conserve their total 
number is to add more electrons to the $\pi$ bands, which thus come 
into play.  To obtain this, the Fermi level needs to shift, and the 
dependence on $u$ of the total energy, is, in turn, deeply 
modified:
\begin{equation}
\mu(u)=\frac{N_\sigma}{N_\sigma+N_\pi} \left(g|u|-
\epsilon_\sigma^{\rm top}\right)
\hspace{8mm}
g |u| \ge \epsilon_\sigma^{\rm top}.
\label{musmall}
\end{equation}
\begin{eqnarray}
E(u)&=&E(0)+\frac{M\omega_{2g}^2}{2}u^2 - 2 N_\sigma g^2 u^2 
\label{enlarge}
\\ 
&&+
\frac{N_\sigma(2N_\sigma +N_\pi)}{N_\sigma + N_\pi} (g|u|-
\epsilon_\sigma^{\rm top})^2
\hspace{4mm}
g |u| \ge \epsilon_\sigma^{\rm top}.\nonumber
\end{eqnarray}
In our simple model the transition between harmonic and anharmonic
regime occurs when one band is completely shifted below the Fermi level,
and does not manifest itself as a simple additional quartic term $u^4$. 
Rather, an overall anharmonic potential results from a simple harmonic
term up to $g|u| \le \epsilon_\sigma^{\rm top}$ which, for $g |u| \ge
\epsilon_\sigma^{\rm top}$, smoothly connects to a shifted parabola with
different curvature.  Such a non--analytic behaviour has to do with the
extreme simplifications of our model, in particular with the perfectly
2D parabolic character of the $\sigma$ bands (step--like density of
states), and on the assumption of perfect degeneracy of the $\sigma$
bands at $u\!=\!0$ [$\epsilon_1({\bf k})\!=\!\epsilon_2({\bf k})$].  We
have checked that, with slightly more realistic models, the sharp
transition of Eq.~(\ref{enlarge}) becomes considerably smoother. 
However, Eq.  (\ref{enlarge}) is particularly appealing just because of
its simplicity, since it depends only on a few parameters which can be
extracted from LDA calculations (see Table~\ref{table1}), thus providing
a direct test of the model.  The results are shown as solid lines in
Fig.~\ref{etot} for the compounds considered here.  The agreement with
LDA first-principles calculations is quite good, considering the extreme
simplifications of our model.  In conclusion, in this paper we have
investigated, by means of numerical and analytical techniques, the
microscopic nature of the anharmonicity of the $E_{2g}$ phonon mode in
MgB$_2$.  The results presented here provide a clear evidence that the
anharmonicity of the $E_{2g}$ phonon mode in MgB$_2$ is induced by the
extremely small value of the $\sigma$--hole Fermi energy.  Along this
view we can predict a strong anharmonicity in heavily hole--doped
graphite.  We have shown that the anharmonicity of the$E_{2g}$ mode,
which is strongly coupled to the $\sigma$ bands, can be considered a
signature of small Fermi energy; this points out in a natural way
towards the possibility of nonadiabatic effects.  A quantitative
description of this situation involves, however, quantum-many-body effects
(nonadiabatic renormalization of the phonon
frequencies\cite{alexandrov1}). While this task is beyond the aim of
the present paper, different theoretical studies already suggest that
nonadiabatic effects could be responsible for the high $T_c$ in
MgB$_2$\cite{alexandrov2,ccgps}.  Note that this is entirely different
from the initial claim \cite{yildirim}, that anharmonicity affects
superconductivity via the nonlinear coupling.  From this point of view
the situation is, instead, similar to fullerenes \cite{cgps}, which have
been recently shown to reach critical temperatures as high as $T_c=117$K
in FET doped compounds \cite{schoen117}.  In this respect our work
suggests new perspectives in the search for high--$T_c$ materials.  In
particular, heavily hole--doped graphite, which we predict to have small
Fermi energy and anharmonic $E_{2g}$ phonon, would be a potential
candidate for high-- $Tc$ superconductivity.  The recent claim of
$T_c=35$K in amorphous graphite--sulfur composite samples could be
related to this scenario \cite{dasilva}.


\begin{thebibliography}{100}
\bibitem{akimitsu}
J. Nagamatsu, N. Nakagawa, T. Muranaka, Y. Zenitani, J. Akimitsu,
Nature {\bf 410}, 63 (2001).
\bibitem{kong}
Y. Kong, O.V. Dolgov, O. Jepsen, and O.K. Andersen,
Phys. Rev. B {\bf 64},020501  (2001).
\bibitem{bohnen}
K.-P. Bohnen, R. Heid, and B. Renker, 
Phys. Rev. Lett. {\bf 86}, 
5771 (2001).
\bibitem{hirsch}
J.E. Hirsch, Phys. Lett. A {\bf 282}, 392 
(2001).
\bibitem{yildirim}
T. Yildirim, O. Gulseren, J.F. Lynn, C.M. Brown, T.J. Udovic, 
Q. Huang, N. Rogado, . A. Regan, M.A. Hayward, J.S. Slusky,
T. He, M.K. Haas, P. Khalifah, K. Inumaru, and R.J. Cava,
Phys. Rev. Lett. {\bf 87}, 037001 (2001).
\bibitem{kunc}
K. Kunc, I. Loa, K. Syassen, R.K. Kremer, K. Ahn,
J. Phys.: Cond. Mat. {\bf 13}, 9945 (2001).
\bibitem{liu}
A.Y. Liu, I.I. Mazin, and J. Kortus,
Phys. Rev. Lett. {\bf 87}, 087005 
(2001).
\bibitem{choi}
H. J. Choi, D. Roundy, H. Sun, M. L. Cohen, S. G. Louie,
cond-mat/0111182 (2001).
\bibitem{struzhkin}
V.V. Struzhkin, A.F. Goncharov, R.J. Hemley, H. K. Mao, G. Lapertot, 
S. L. Bud'ko, P. C. Canfield,
cond-mat/0106576 (2001).
\bibitem{meletov} This is quite different from
what discussed in
K.P. Meletov, J. Arvanitidis, M.P. Kulakov, N.N. Kolesnikov, G.A. Kourouklis,
cond-mat/0110511 (2001), where a $\sigma$-band 
splitting was assumed even in the undistorted lattice.
\bibitem{kortus}
J. Kortus, I.I. Mazin, K.D. Belashchenko, V.P. Antropov, L.L. Boyer,
Phys. Rev. Lett. {\bf 86}, 4656 (2001).
\bibitem{Satta}
G. Satta, G. Profeta, F. Bernardini, A. Continenza, S. Massidda,
Phys. Rev. B {\bf 64}, 104507 
(2001).
\bibitem{Pickett}
J.M. An, W.E. Pickett,
Phys. Rev. Lett. {\bf 86}, 4366 (2001).
\bibitem{hinks}
D.G. Hinks, H. Claus, J.D. Jorgensen,
Nature {\bf 411}, 457 (2001).
\bibitem{dresselhaus}
M.S. Dresselhaus and G. Dresselhaus,
Adv. Phys. {\bf 30}, 139 (1981).
\bibitem{HKS} P. Hohenberg e W. Kohn, Phys.  Rev.  {\bf 136}, 
B864 (1964); W. Kohn e L.J. Sham, Phys.  Rev.  {\bf 140}, A1133 
(1965). 
\bibitem{MartTroul}
N. Troullier, J. L. Martins, Phys. Rev. B {\bf 43}, 1993 (1991). 
\bibitem{ABINIT} The ABINIT code is a common project
of the Universit\'e Catholique de Louvain, Corning Incorporated, 
and other contributors (URL http://www.abinit.org). 
\bibitem{alexandrov1}
A.S. Alexandrov and H. Capellmann,
Phys. Rev. B {\bf 43}, 2042 (1991).
\bibitem{alexandrov2}
A.S. Alexandrov, Physica C {\bf 363}, 231 
(2001).
\bibitem{ccgps}
E. Cappelluti, S. Ciuchi, C. Grimaldi, L. Pietronero, 
and S. Str\"assler
Phys. Rev. Lett. {\bf 88}, 117003 (2002).
\bibitem{cgps}
E. Cappelluti, C. Grimaldi, L. Pietronero, 
and S. Str\"assler,
Phys. Rev. Lett. {\bf 85}, 4771 (2000).
\bibitem{schoen117}
J.H. Sch\"on, Ch. Kloc, and B. Batlogg,
Science {\bf 293}, 2432 (2001).
\bibitem{dasilva}
R. Ricardo da Silva, J. H. S. Torres, and Y. Kopelevich,
Phys. Rev. Lett. {\bf 87}, 147001 (2001).
\end{thebibliography}
\end{document}